\RequirePackage{xcolor}
\documentclass[journal,twoside,web]{ieeecolor}
\usepackage{etoolbox}
\makeatletter
\@ifundefined{color@begingroup}%
  {\let\color@begingroup\relax
   \let\color@endgroup\relax}{}%
\def\fix@ieeecolor@hbox#1{%
  \hbox{\color@begingroup#1\color@endgroup}}
\patchcmd\@makecaption{\hbox}{\fix@ieeecolor@hbox}{}{\FAILED}
\patchcmd\@makecaption{\hbox}{\fix@ieeecolor@hbox}{}{\FAILED}

\usepackage{tmi}
\usepackage{cite}
\usepackage{amsmath,amssymb,amsfonts}
\usepackage{algorithmic}
\usepackage{graphicx}
\usepackage{textcomp}
\usepackage{url}
\def\BibTeX{{\rm B\kern-.05em{\sc i\kern-.025em b}\kern-.08em
    T\kern-.1667em\lower.7ex\hbox{E}\kern-.125emX}}

\begin{document}

\title{Motion simulation of radio-labeled cells in whole-body positron emission tomography}

\author{Nils Marquardt, Tobias Hengsbach, Marco Mauritz, Benedikt Wirth, and Klaus Schäfers
\thanks{This work has been accepted for publication by the IEEE under DOI 10.1109/TMI.2025.3614767. © 2025 IEEE. Personal use of this material is permitted.  Permission from IEEE must be obtained for all other uses, in any current or future media, including reprinting/republishing this material for advertising or promotional purposes, creating new collective works, for resale or redistribution to servers or lists, or reuse of any copyrighted component of this work in other works.}
\thanks{This work was funded by the Deutsche Forschungsgemeinschaft (DFG) – CRC 1450 – 431460824.}
\thanks{N. Marquardt, T. Hengsbach and K. Schäfers are with the
European Institute for Molecular Imaging, University of Münster, 48149 Münster, Germany (e-mail: n.marquardt@uni-muenster.de; thengsba@uni-muenster.de; schafkl@uni-muenster.de).}
\thanks{M. Mauritz and B. Wirth are with the
Institute for Computational and Applied Mathematics, University of Münster, 48149 Münster, Germany (e-mail: marco.mauritz@uni-muenster.de; benedikt.wirth@uni-muenster.de).}}

\maketitle
\bstctlcite{IEEEexample:BSTcontrol} 
\begin{abstract}
Cell tracking is a subject of active research gathering great interest in medicine and biology. Positron emission tomography (PET) is well suited for tracking radio-labeled cells in vivo due to its exceptional sensitivity and whole-body capability. For validation, ground-truth data are desirable that realistically mimic the flow of cells in a clinical situation. This study develops a workflow (CeFloPS) for simulating moving radio-labeled cells in a human phantom. From the XCAT phantom, the blood vessels are reduced to nodal networks along which cells can move and distribute to organs and tissues. The movement is directed by the blood flow, which is calculated in each node using the Hagen-Poiseuille equation and Kirchhoff’s laws assuming laminar flow. Organs are voxelized and movement of cells from artery entry to vein exit is generated via a biased 3D random walk. The probabilities of cells moving or remaining in tissues are derived from rate constants of tracer kinetic-based compartment modeling. PET listmode data is generated using the Monte-Carlo simulation framework GATE based on the definition of a large-body PET scanner with cell paths as moving radioactive sources and the XCAT phantom providing attenuation data. From the flow simulation of 100,000~cells, 100 sample cells were further processed by GATE and listmode data was reconstructed into images for comparison. As demonstrated by comparisons of simulated and reconstructed cell distributions, CeFloPS is capable of simulating cell behavior in whole-body PET. It achieves this simulation in a way that is anatomically and physiologically reasonable, thereby providing valuable data for the development and validation of cell tracking algorithms.
\end{abstract}

\section{Introduction}
\label{sec:introduction}
\IEEEPARstart{C}{ell imaging} and tracking are subjects of active research gathering great interest in the medical and biomedical science community. Non-invasive imaging techniques using fluorescent or luminescent markers in optical imaging (OI), iron oxide particles in magnetic resonance imaging (MRI), or radio-labeled tracers in single photon emission computed tomography (SPECT) and positron emission tomography (PET) have been frequently used in preclinical studies to permit in vivo visualization of living cells \cite{Kircher2011}.
For human application, MRI, PET, and SPECT are the most promising tools as they provide either high spatial resolution (MRI) or excellent whole-body sensitivity (PET, SPECT). Of these imaging modalities, PET has become of special interest following labeled immune cells in the context of stem cell imaging and cell-based therapies \cite{Ponomarev2017}. 

PET is particularly advantageous because it combines exceptional sensitivity with the possibility of whole-body coverage, which has been further advanced in recent years by the design of large-field-of-view (LAFOV) PET systems with extended coverage of the axial field-of-view in the range of 1 to 2 meters, improved time-of-flight capabilities and excellent system sensitivities \cite{Cherry2018}. Using either long-lived isotopes or reporter genes, immunoPET has great potential to identify biomarkers allowing patients to be treated based on their most likely response to a dedicated immunotherapy \cite{Friberger2023, Jung2019, Melendez2023}. 

The idea of imaging or tracking biological cells with PET is highly related to the field of Positron Emission Particle Tracking (PEPT) which has been developed primarily for technical applications, such as evaluation of particle–fluid or particle–particle interactions in the pharmaceutical, chemical, oil, mining, construction and power generation industries \cite{windows2022}. The idea is to follow one or a few radio-labeled particles by tracking the coincidence lines-of-response (LORs) recognized from its two back-to-back gamma rays emitted after beta-plus decay. In contrast to PET, where volumetric images of the radioactivity distribution are reconstructed from LORs, PEPT algorithms calculate a trajectory by connecting LORs in the most probable way, offering direct motion tracking of radio-label particles \cite{pellico2024vivo}. 
As PEPT applications are generally carried out outside a clinical environment, special industrially usable devices were developed in parallel to the invention of clinical PET systems \cite{Hawkesworth1986, Hampel2022}. 
Following the idea of PEPT in a clinical or pre-clinical environment, the outstanding sensitivity of PET can be utilized to track radio-labeled singular cells traveling inside a biological system (e.g. human, mouse) \cite{Hampel2022, Ouyang2016a, Ghosh2023, Nguyen2024}. 

The here presented work describes two simulations that build a foundation for development and validation of cell tracking in humans. The first simulation mimics the movement of radio-labeled biological cells in blood and their accumulation in tissues of the human body. To this end, the XCAT phantom was used to define the hemodynamic parameters for vascular blood flow, accompanied by a random walk approach to mimic the cell motion in the vascular bed \cite{Segars2010_XCAT}. Transition probabilities for the accumulation of cells in tissue were implemented via a tracer kinetic model approach. An XCAT-based approach has been used before to simulate blood flow in the vascular tree \cite{segars2017application, 10.1145/2807591.2807676}. The simulation of the transition of cells into the tissue and back via tracer kinetic models represents a new extension by us. Tracer kinetic models are commonly utilized in dynamic PET studies to measure how substances are distributed, metabolized, and eliminated within the body \cite{kotasidis2014advanced, lammertsma2017forward}. Vice versa, these model parameters can also be utilized to simulate cell transitions. 

The second simulation enables the generation of raw PET data from the simulated cell pathways using Monte Carlo based physics simulations. Such simulations were also used in the PEPT context based on the Geant4 Application for Tomographic Emission (Gate) \cite{Jan2004_GATE}. For example, Herald et al. simulated both the ADAC/Phillips Forte and the Siemens Inveon pre-clinical PET and compared fluid simulations with experimental data \cite{Herald2018, Herald2021}. The foundation of our Gate simulation, developed here, is the concept of new total-body PET with an extensive field of view and high sensitivity, which presents novel possibilities for human whole-body cell tracking.

The task of cell tracking is not part of this work, but has been developed in many variants by different groups in the past \cite{Parker1993, Jung2019}. Nguyen et al. recently presented a new algorithm, PEPT-EM, which was able to successfully track the fate of more than 70 radio-labeled melanoma cells in a mouse model \cite{Nguyen2024}. In this context, the group recently introduced an image reconstruction algorithm for PET that employs an optimal transport-based regularisation technique. This algorithm demonstrates considerable potential for the tracking of radioactively labeled cells within a human body, with no limitation on the number of trackable cells, and with the capacity to address the inherent handling of scattered, random, and background events. The proposed methodology integrates the information from all detected events to facilitate the reconstruction of dynamic evolution, while concurrently enhancing the reconstruction quality through the enforcement of temporal consistency. \cite{Schmitzer2019_OptimalTransport, mauritz2024, Mauritz_2025}

This study provides a unique workflow for cell flow PET simulation (CeFloPS) to simulate the movement of radio-labeled cells within the human body, serving as ground-truth data for developing and validating novel image reconstruction and cell tracking tools.

\section{Methods}\label{sec3}

\subsection{Simulation basis}\label{subsec3.1}

\begin{figure}[ht]
\centerline{\includegraphics[width=\columnwidth]{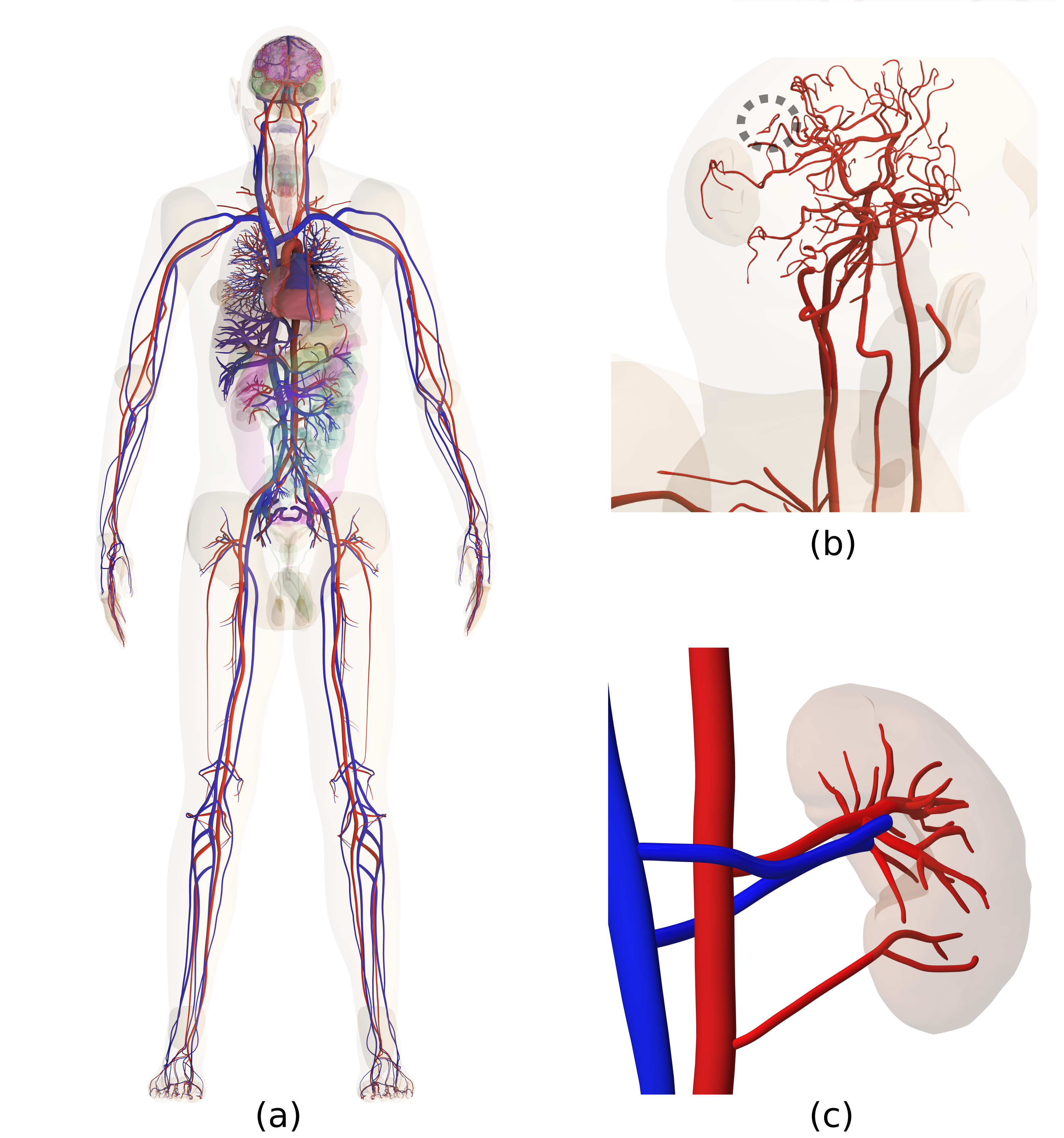}}
\caption{Mesh representation of the XCAT phantom defining vessel structures and organs (a). Only the larger vessels are defined by XCAT showing partly missing connections, exemplarily shown on the arteries in the brain (b) and a just coarsely defined capillary bed, here shown for the kidney (c).}
\label{XCAT}
\end{figure}

CeFloPS is based on anatomical data provided by the XCAT phantom with high-resolution 3D models of all organs, blood vessels, bones, and muscles (Fig.~\ref{XCAT}a) in the form of STEP files \cite{ISO10303}. Using Inventor (Version 2023, Autodesk Inc., San Rafael, USA), these STEP files were converted into StL files (StereoLithography file format), yielding triangulated surface meshes of organs and structures. The StL files were further processed in Python (Version 3.11.5, Python Software Foundation) with the help of the trimesh library \cite{trimesh}, allowing for operations on the meshes and giving access to the coordinates of the vertices. For further processing, meshes of muscles and internal organs were voxelized at a resolution of 0.5~mm~$\times$~0.5~mm~$\times$~0.5~mm, whereas the complex anatomy of blood vessel meshes was simplified by reducing it to node networks. The nodes describe the possible paths a cell can take and are placed along the central axis of a blood vessel, storing information about (i) its spatial coordinates, (ii) which nodes are adjacent in the direction of blood flow and (iii) the radius of the blood vessel. 

\begin{figure}[ht]
\centerline{\includegraphics[width=.75\columnwidth]{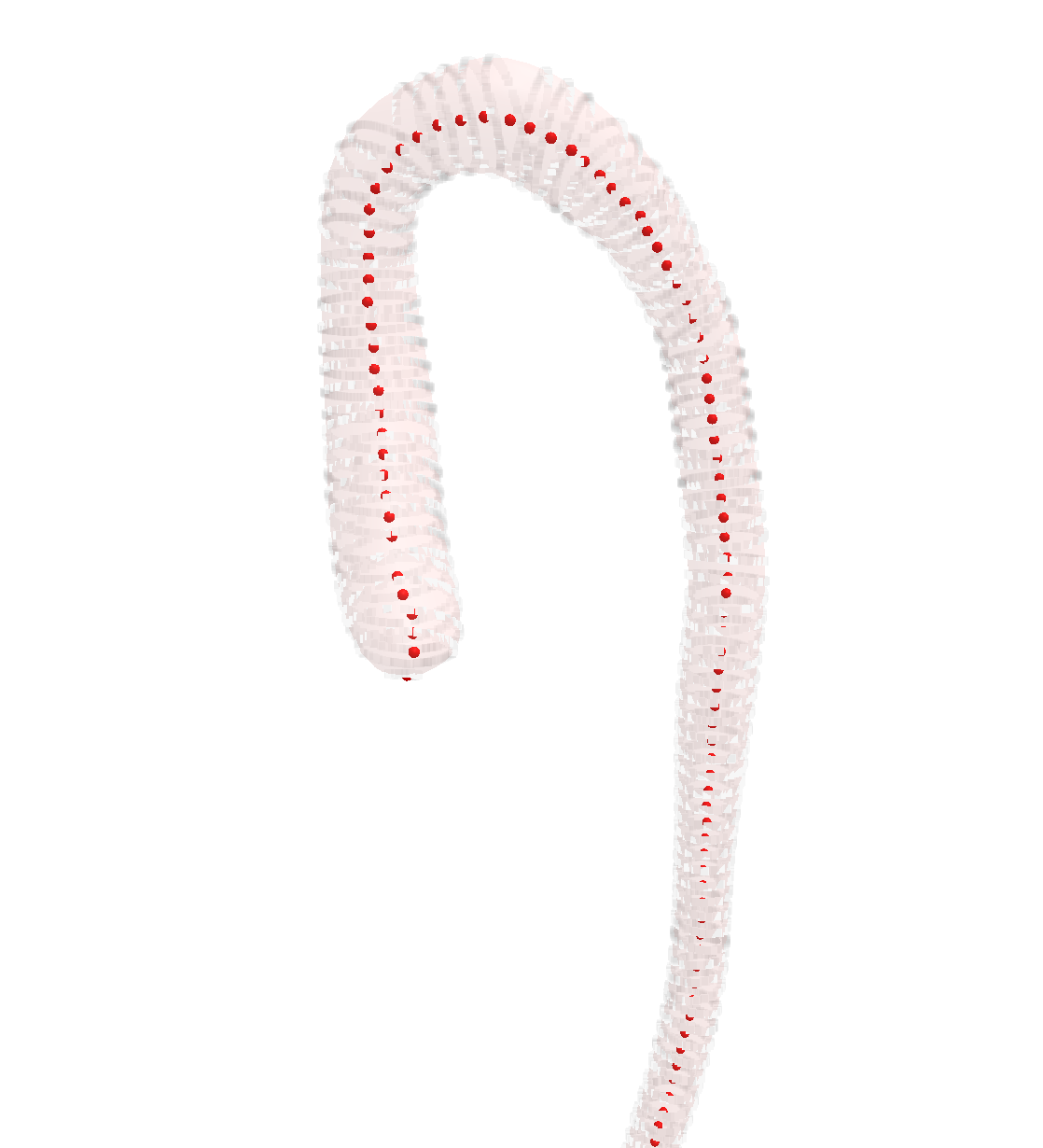}}
\caption{XCAT based mesh of the aorta together with the node representation (red dots) calculated from orthogonal planes. Each node stores information about next node neighbors, blood flow direction, and vessel radius.}
\label{aorta_nodes}
\end{figure}

To generate the nodes, the blood vessels were divided into segments starting from one end point of a vessel. For each segment, the barycenter was formed of all vertices located in this segment, resulting in points lying inside a blood vessel. Starting from the first point inside a blood vessel, planes, orthogonal to the blood vessel, were then placed at an interval of 1~mm along the vectors between these points. Using the intersection of those planes with the blood vessel surface, the vessel radius and the new center point were calculated, resulting in the final node positions spaced at 1~mm (see Fig.~\ref{aorta_nodes}). 

Nodes that lie at bifurcations of blood vessels and thus have more than one neighbor in the direction of blood flow were marked and are hereafter referred to as links. Since the spatial resolution of the XCAT phantom is limited, some connections between vessels are missing, as seen in Fig.~\ref{XCAT}b. In such cases, we interpolated linearly between the nodes of those vessels to generate a connection. In addition, the spatial resolution of blood vessels varies from organ to organ. The smallest radius defined in XCAT is 0.15 mm and is located in the toes. Consequently, no capillary bed is included that connects the nodes of arteries and veins. (Fig.~\ref{XCAT}c). This results in a total of five separate networks of nodes: arteries, veins, portal vein, and pulmonary arteries and veins. All of these, except the portal vein, are connected via the heart. In order to compensate for the lack of capillaries, these separate networks are artificially combined to form a single closed network in which opposing arteries and veins leading to the same organ are virtually connected.


\subsection{Blood flow simulation}\label{subsubsec3.2}

From the node networks, the mean blood flow rate $\overline{Q}_i$ and the vessel resistance $R_i$ were estimated for each node $i$ from a highly simplified flow model using Kirchhoff’s laws in combination with the Hagen-Poiseuille equation. This was done by assuming laminar flow in every vessel, each treated as a cylinder \cite{Lampe2014, Helthuis2020}: 
\begin{equation}\label{eq:average_Q}
\overline{Q}_i = \frac{\mathit{\Delta} P_i}{R_i}\,, 
\end{equation}
\begin{equation}\label{eq:vessel_resistance}
R_i = \frac{8\eta l_i}{\pi r^4_i}\,,
\end{equation}
\begin{equation}\label{eq:serial_connection}
R_{\mathrm{tot}} = \sum_{i=1}^n R_i 
\end{equation}
for serial connections of vessels and
\begin{equation}\label{eq:parallel_connection}
\frac{1}{R_{\mathrm{tot}}} = \sum_{i=1}^n \frac{1}{R_i}
\end{equation}
for parallel connections, with $\mathit{\Delta} P_i$ the blood pressure difference, $l_i$ the distance between two nodes, $r_i$ the radius of the vessel at the node position, $n$ the number of nodes in a branch and $\eta$ the dynamic viscosity of blood. The calculations were performed under the assumption of a cardiac output of 100~ml/s, a mean arterial blood pressure of 100~mmHg in the part of the aorta in close proximity to the heart, and 0~mmHg in the vena cava entering the heart. Furthermore, a constant blood viscosity of 3.5~mPa$\cdot$s was presumed. The assumption of laminar flow in the vessels was validated by calculating the Reynolds number 
\begin{equation}\label{eq:Reynolds}
\mathrm{Re}=\frac{2 \rho \overline{Q}_{i}}{\pi r_i \eta}\,,
\end{equation}
with $\rho$ the density of blood set to 1.06~g/ml. Laminar flow is assumed valid if the Reynolds number is below 2000, which is the known threshold beyond which laminar flow gradually transitions into turbulent flow (Reynolds number \textgreater~4000) \cite{CDT9829}. Due to the lack of small blood vessels in the XCAT phantom, our flow model misses a large portion of high resistances which would add to the calculation of the total resistance $R_\mathrm{tot}$ in \eqref{eq:serial_connection} and \eqref{eq:parallel_connection}. By \eqref{eq:average_Q}, this also affects the blood flow $\overline{Q}_i$ in each node.

At a link, the probability $p_\mathrm{node}$ with which a cell will move to the next possible node is calculated based on the blood flow $\overline{Q}_\mathrm{link}$ in the link and $\overline{Q}_{i}$ in each following node: 
\begin{equation}\label{eq:link_probability}
    p_{\mathrm{node},i}=\frac{\overline{Q}_{i}}{\overline{Q}_\mathrm{link}}\quad\mathrm{with}\quad\overline{Q}_\mathrm{link}=\sum_{i=1}^n \overline{Q}_{i}
\end{equation}
and $n$ the number of following nodes of a link.

The velocity of an individual cell is assigned by pseudo-random number sampling based on a cumulative distribution function (CDF). This ensures that a group of cells starting from the same node have different velocities and disperse over time in the blood vessels. The CDF was derived using the velocity profile of laminar flow in a tube \cite{Belhocine2016}
\begin{equation}\label{eq:velocity_profile}
    v(x)=v_\mathrm{max}\left(1-\frac{x^2}{r^2}\right)\quad\mathrm{with}\quad v_\mathrm{max}=\frac{r^2\mathit{\Delta} P }{4\eta l}=\frac{2\overline{Q}}{\pi r^2}\,,
\end{equation}
where $x\in\left[0,r\right]$ is the radial distance from the center. The distribution of blood flow over the profile of the vessels is then given by
\begin{equation}\label{eq:dQ/dx}
    \biggl|\frac{\mathrm{d}Q(x)}{\mathrm{d}x}\biggr|=\pi r^2\biggl|\frac{\mathrm{d}v(x)}{\mathrm{d}x}\biggr|=2Q_{\mathrm{max}}\frac{x}{r^2}\,,
\end{equation}
with $Q_{\mathrm{max}}=2\overline{Q}$. Division by the maximum blood flow rate yields the probability density function (PDF)
\begin{equation}\label{eq:f_P}    
    f_p(x)=\frac{1}{Q_\mathrm{max}}\biggl|\frac{\mathrm{d}Q(x)}{\mathrm{d}x}\biggr|=2\frac{x}{r^2}\ \mathrm{with}\ \int_0^r f_p(x)\mathrm{d}x=1\,,  
\end{equation}
which by integration leads to the CDF:
\begin{equation}\label{eq:F_P}
    F_p(x)=\int_0^x f_p(t)\mathrm{d}t=\frac{x^2}{r^2}\,.
\end{equation}
The velocity of a cell is then obtained by \eqref{eq:velocity_profile} with a randomly picked $x$ according to \eqref{eq:F_P}.

\subsection{Organ distribution simulation}\label{sec4}

Inside a voxelized organ, cell motion is guided by a biased 3D random walk along the voxels to simulate the capillary bed and the tissue exchange. This process is carried out in a series of discrete calculations, each with a defined time step. The current simulations have set the time step to 600 ms. First, voxels containing an artery entrance or a vein exit were identified and hereafter referred to as start-voxels and end-voxels, respectively. If an organ does not have an artery entrance and/or vein exit due to missing connections in the XCAT phantom, the nearest artery/vein is calculated and the closest voxel to that blood vessel is determined as start-/end-voxel. 

\begin{figure}[ht]
\centerline{\includegraphics[width=\columnwidth]{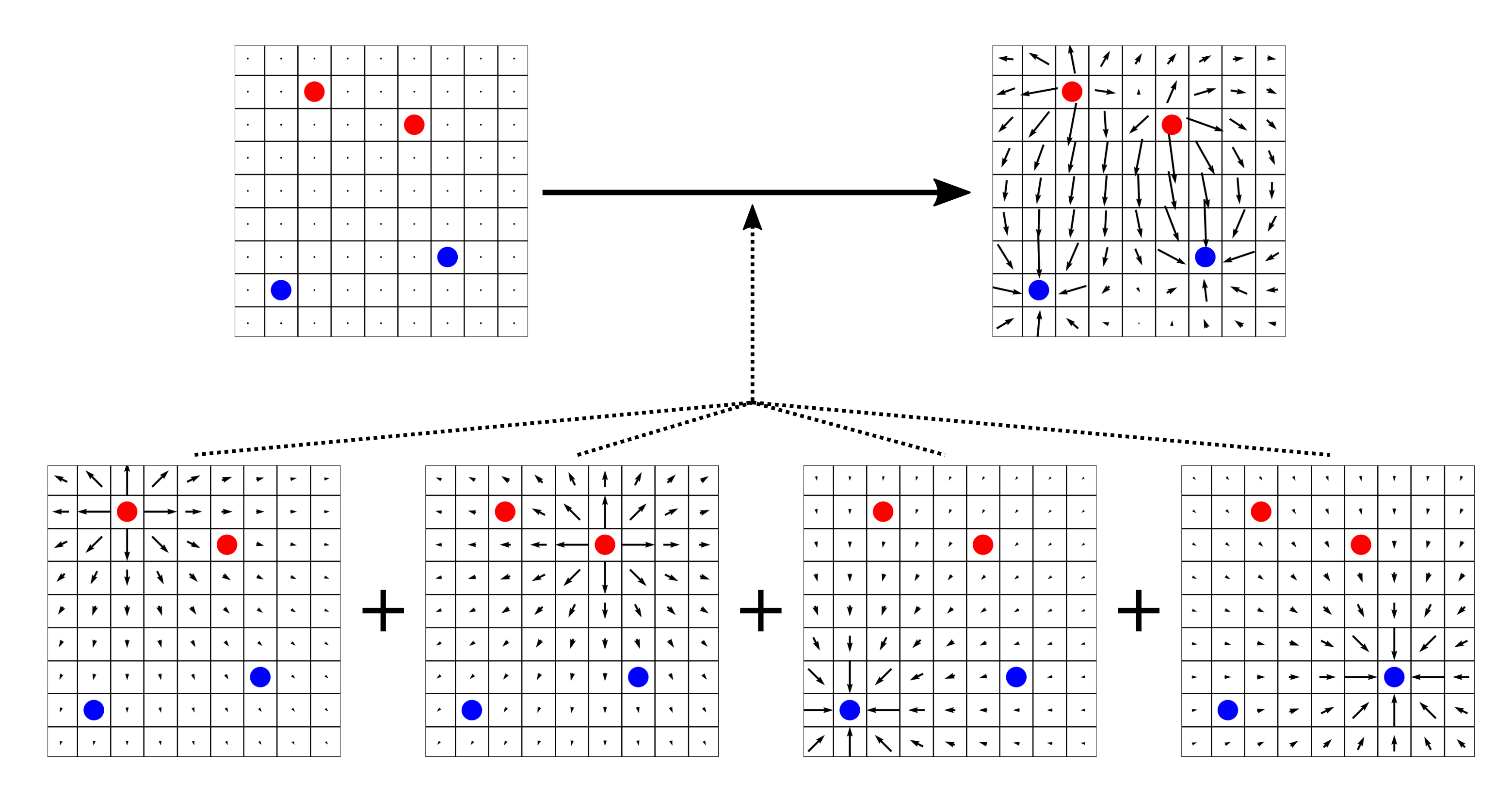}}
\caption{Principle of a biased random walk for a single cell defined by arterial sources (red) and venous sinks (blue) (top left). For each voxel, the most probable direction vector (top right) is calculated by superimposing attracting and repellent vectors for each start- and end-point (bottom).}
\label{2D_random_walk}
\end{figure}

Remaining blood vessels are connected to an organ in the same way if their distance to that organ lies within a 2~cm threshold. To compensate for missing arterial branches in muscles and organs, each node of the arteries and veins can serve as entrance and exit for cells, respectively. For every voxel within an organ or muscle, direction vectors towards all start- and end-voxels were determined along with separate weighting values. As the distance $d$ between a voxel and its start/end voxel increases, these weightings decrease by a factor of $1/d^2$, as shown in Fig.~\ref{2D_random_walk}. The vectors point away from start-voxel or towards end-voxel. The combination of those vectors represents the most probable directions for a cell to move. The weightings of vectors in start- and end-voxels were set to zero. 

The random walk is initiated at a start-voxel, where a cell can move to one of the surrounding voxels. This yields 26 possible directions (provided the start-voxel does not reside on the organ's periphery), all with the same probability. For the proceeding steps, the direction is biased in accordance with the previously determined weightings of the direction vectors in each voxel. If a cell reaches an end-voxel, the random walk concludes, followed by the cell exiting via the vein back into the nodal network for further blood flow simulation. The speed at which cells perform the random walk in the different organs was calibrated against the delay times determined by \cite{leggett1995proposed}. For this purpose, first the distribution of required time to travel from start- to end-voxel in each organ is determined for cells at an arbitrary random walk speed. The mean value of this distribution is calculated and compared with the value derived by \cite{leggett1995proposed}, leading to new speed values for the corresponding organ.

\begin{figure}[ht]
\centerline{\includegraphics[width=\columnwidth]{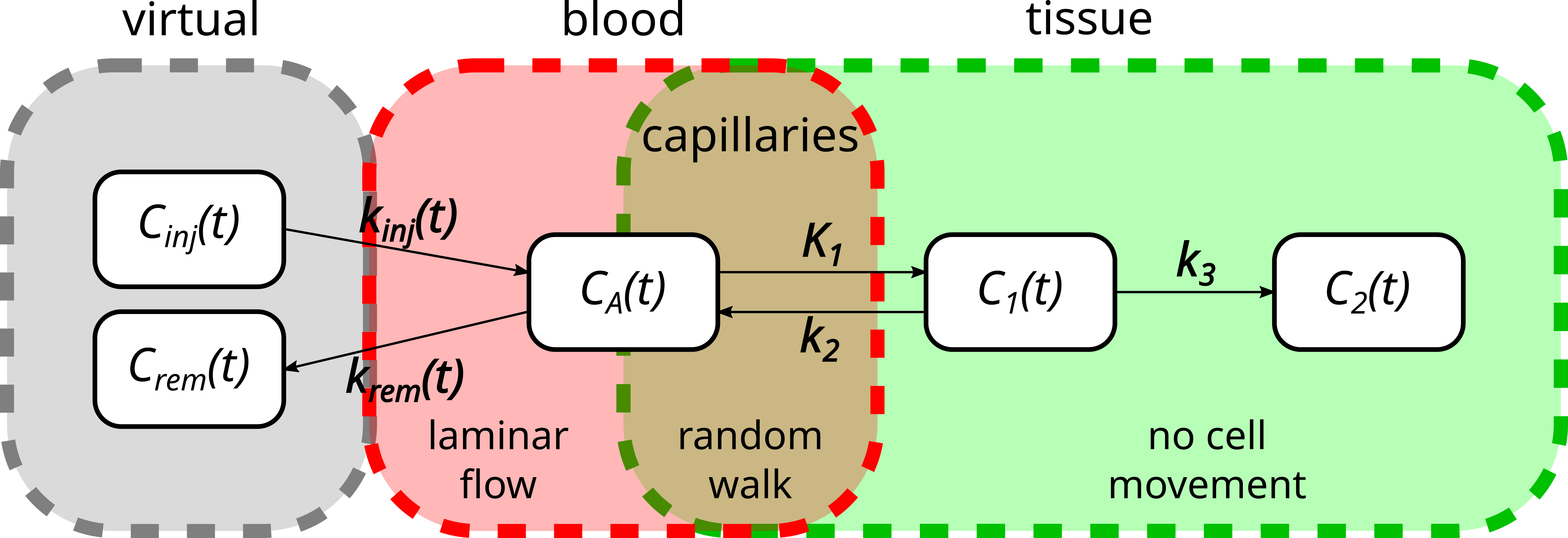}}
\caption{The movement of cells within each tissue is regulated by compartmental models, which determine the probability of a cell undergoing a random walk or being constrained to a compartment.}
\label{Compartments}
\end{figure}

In addition to the biased random walk, a compartment model was incorporated into the organ distribution simulation to account for the transition state of a cell in tissue. Besides its spatial voxel location, each cell occupies a compartmental state, which dictates whether a cell can move or not. The exchange between compartments is determined by transition rates $k_i$. The number of compartments and the values of the transition rates must be known in advance to the CeFloPS simulation and can be obtained from (physiologically based) kinetic models. As a first example and as proof-of-principle, we employ a two-compartment model comprising the compartments $C_1$ and $C_2$ (shown in Fig.~\ref{Compartments}). The compartment $C_1$ signifies a reversible binding of the cells, whereas $C_2$ denotes an irreversible binding. Consequently, if a cell is situated within either $C_1$ or $C_2$, it does not execute a random walk and remains temporarily or permanently in the same voxel. The change in amount of cells in these compartments can be described with the transition rates $K_1$, $k_2$, and $k_3$ through the differential equations 
\begin{align}
    \begin{split}
        \frac{\mathrm{d}}{\mathrm{d}t}C_1(t) &= K_1C_A(t)-(k_2+k_3)C_1(t)\quad\mathrm{and}\\
        \frac{\mathrm{d}}{\mathrm{d}t}C_2(t) &= k_3C_1(t)\,.
    \end{split}
    \label{diffequations_two}
\end{align}
The arterial input function (AIF), denoted by $C_A(t)$, serves as a metric for the concentration or number of cells present in the arterial blood and also has to be entered as input into the CeFloPS simulation. One method for generating an AIF is outlined in \cite{FENG}. It is important to note that, in general, transition rates are only defined in the underlying kinetic model for a subset of the tissues included in CeFloPS. Consequently, the sum of all cells transitioning from the arterial blood into the specified tissues may not be equal to the rate of change of the AIF. In CeFlops, this would result in an erroneous distribution of cells, which would not correspond to \eqref{diffequations_two}. To counteract this, two virtual compartments were introduced. The first compartment, denoted by $C_\mathrm{inj}$, represents the injection of cells into the arterial blood during the beginning of the simulation. The second compartment, designated by $C_\mathrm{rem}$, describes cells which are removed from the arterial blood and go into a tissue not defined in the kinetic model rates. The rate of change of the AIF is then given by 
\begin{align}
    \begin{split}
    \frac{\mathrm{d}}{\mathrm{d}t}C_A(t) &= k_\mathrm{inj}(t)C_\mathrm{inj}(t)-k_\mathrm{rem}(t)C_A(t)\\&+\sum_\mathrm{tissues}\left[ k_2C_1(t)-K_1C_A(t)\right]\,.
    \end{split}
    \label{AIF_equation}
\end{align}
The transition rates $k_\mathrm{inj}$ and $k_\mathrm{rem}$ are selected as time-dependent to account for the time-limited injection phase of the cells at the beginning of a simulation and potential delays during the transition into tissues. The determination of $k_\mathrm{inj}(t)$ and $k_\mathrm{rem}(t)$ is achieved through the implementation of a 3-layer neural network. The neural network identifies the optimal values for $k_\mathrm{inj}$ and $k_\mathrm{rem}$ at each simulation time $t$ by fitting the cell distributions in all tissues to fulfill the differential equations from \eqref{diffequations_two} and \eqref{AIF_equation}. The latter phase of the distributions, in which a state of equilibrium is ultimately achieved, is given greater weight relative to the initial phase, during which significant alterations may occur. This ensures that the equilibrium phase at late stages of the simulation is primarily governed by the compartment model. The neural network is set up using the Python library PyTorch \cite{pytorch}.

The transition rates $k_i$ describe the transition behavior exhibited by the entirety of the cell population. However, in the CeFloPS simulation, each cell is considered independently. To determine the transition probability between compartments in a tissue for a single cell, the compartments are treated as states of a continuous Markov chain with a $3\!\times\!3$ transition rate matrix $\mathbf{G}$. The elements $\gamma_{i,j}$ of that matrix denote the transition rate from state $i$ to state $j$. Diagonal elements of $\mathbf{G}$ are determined by the condition that the row-sums of the matrix are equal to 0, i.e. $\sum_{j=1}^3\gamma_{i,j}=0$ for $i=[1,2,3]$. For the two-compartment model with rate constants $K_1$, $k_2$ and $k_3$, this results in the following matrix:
\[ \mathbf{G} = \begin{pmatrix}
-K_1 & K_1 & 0 \\
k_2 & -k_2-k_3 & k_3 \\
0 & 0 & 0 \end{pmatrix}\,.  \]

As shown in \cite{jones2017procedure,welton2005estimation,welton2007solution}, one can derive a matrix $\mathbf{P}(t)$ of transition probabilities from $\mathbf{G}$, with elements $\pi_{i,j}(t)$ giving the transition probability from state $i$ to state $j$ during the length of a time step $t$:
\begin{equation}\label{transition_probabilities}
    \mathbf{P}(t)=\exp(t\mathbf{G})=\sum_{n=0}^{\infty}\frac{t^n}{n!}\mathbf{G}^n\,,
\end{equation}

where $\exp$ is the matrix exponential. It takes into account that more than one transition can occur during a time step. Therefore the probabilities for transitions from $C_A$ to $C_2$ (i.e. $\pi_{1,3}$) are greater than 0, although in the two-compartment model no direct transition can occur between those compartments. In our simulations \eqref{transition_probabilities} is solved numerically using the Python package scipy \cite{2020SciPy-NMeth}.

\begin{figure}[ht!]
\centerline{\includegraphics[width=\columnwidth]{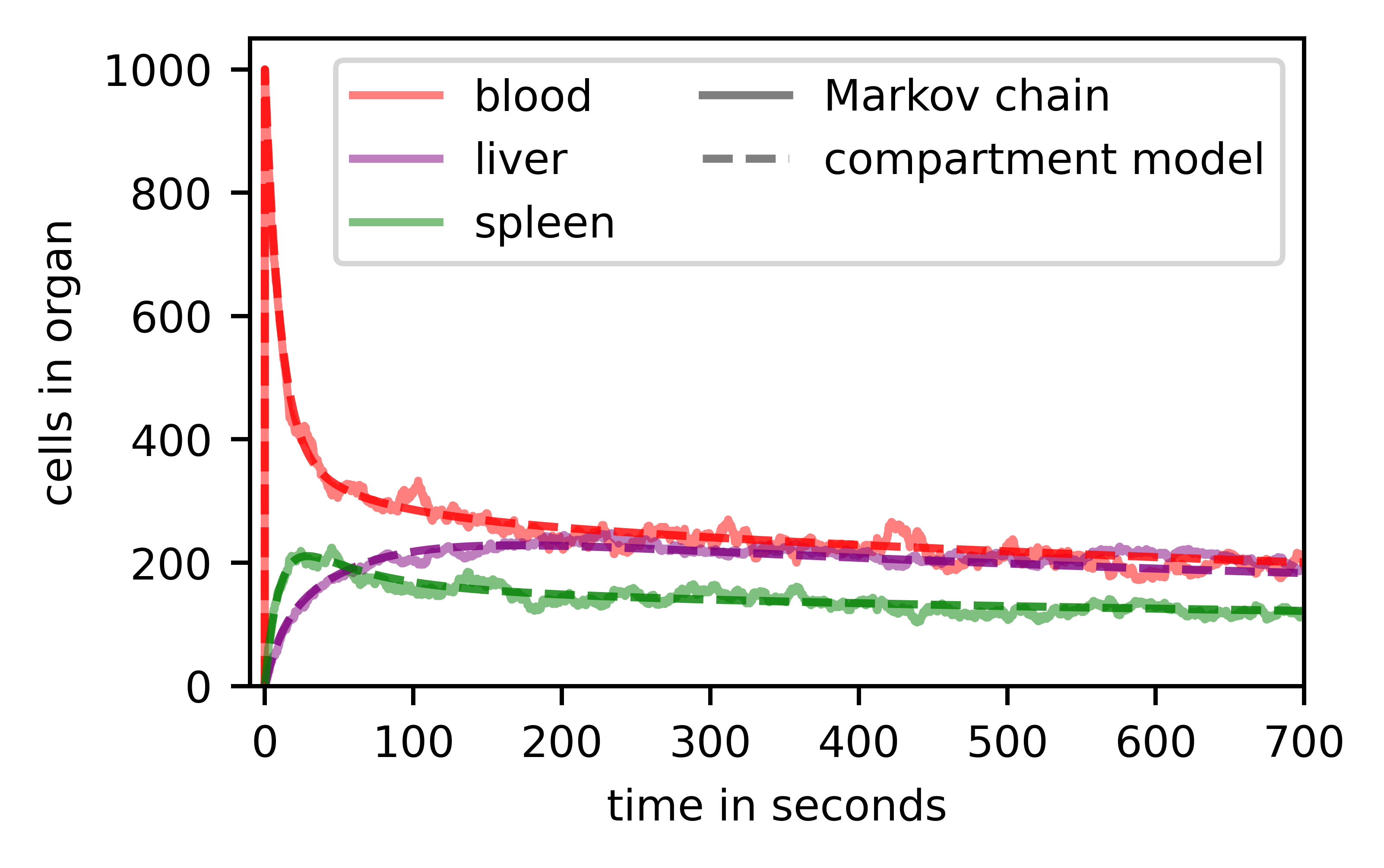}}
\caption{Example model of cell distributions over time in blood, liver, and spleen after an injection at $t=0$ into the bloodstream. The cell dynamics were modeled using a (kinetic) compartment model based on \eqref{diffequations_two} (solid lines) and a Markov chain approach based on \eqref{transition_probabilities} (dashed lines).}
\label{Comp_Markov}
\end{figure}

In this example, we implement the rate constants of a $^{18}$F-Fluordesoxyglucose (FDG) model taken from \cite{Li2022_KineticModelling, kWerte2} and listed in Table~\ref{tab:k_values}. The application of kinetic modeling with FDG has been comprehensively addressed in the extant literature, thereby facilitating a more robust validation of our simulation compared to kinetic modeling with cells, which is still scarce in the literature and usually only takes few tissues into account. The simulated cells in this example therefore do not behave like real, existing cells, but like an FDG tracer. In this context, $C_1$ and $C_2$ denote the unbound and bound states of an FDG "cell" in tissue and $C_A$ the free FDG "cell" in the blood supply of the tissue. Based on the probabilities calculated with \eqref{transition_probabilities} from these transition rates, a random decision is made at each time step as to whether a cell transitions between these compartments. 
Fig.~\ref{Comp_Markov} demonstrates the temporal cell distributions calculated by the Markov chain approach in comparison to the whole population based compartment modelling using differential equations from \eqref{diffequations_two}, showing no systematic differences.

\begin{table}[ht!]
\caption{Values of some rate constants taken from \cite{Li2022_KineticModelling}$^{\mathrm{a}}$ and \cite{kWerte2}$^{\mathrm{b}}$ used in the simulation}
\setlength{\tabcolsep}{3pt}
\centering
\begin{tabular}{|p{50pt}|c|c|c|}
\hline
Tissue & 
\rule[-1.5ex]{0pt}{4ex} $K_1$ [ml/min/ml] &
\rule[-1.5ex]{0pt}{4ex} $k_2$ [min$^{-1}$] &
\rule[-1.5ex]{0pt}{4ex} $k_3$ [min$^{-1}$]\\
\hline
lung$^{\mathrm{a}}$ & 
0.023 &
0.205 &
0.001\\
liver$^{\mathrm{a}}$ & 
0.660 &
0.765 &
0.002\\
spleen$^{\mathrm{a}}$ & 
1.593 &
2.867 &
0.006\\
pancreas$^{\mathrm{b}}$ & 
0.3561 &
1.7077 &
0.0787\\
kidneys$^{\mathrm{b}}$ & 
0.7023 &
1.3542 &
0.1778\\
\hline
\end{tabular}
\label{tab:k_values}
\end{table}

For the FDG simulation, the paths and distributions of 100,000~cells are simulated in parallel. For every time step, the organ and compartment in which each cell is located is recorded. This allows the generation of time profiles of cells in tissues. For a given tissue, the tissue concentration of cells per milliliter or the corresponding cell number (as the sum in $C_1$ and $C_2$) is calculated. The number of cells in the blood is made up of all cells that are still in the node network or in a $C_A$ compartment of an organ.

\subsection{PET simulation}

Raw PET data are generated using the Geant4 application for tomographic emission GATE (version 9.0) \cite{Agostinelli2003_Geant4, Jan2004_GATE}. The cell path outputs from the CeFloPS simulations have the format of GATE placement files and thus are used as generic moving sources in the simulations. The sources, i.e. the cells, are simulated as spheres whose diameters and activities can be specified by the user. For the example shown in this work, a diameter of 10~µm and an activity of 1~kBq per cell were simulated for 100~cells. The activity per cell and number of cells do not reflect a realistic scenario but were chosen solely as a proof-of-principle. The PET scanner used in the simulation is based on the Siemens Biograph Vision Quadra (Siemens Healthineers, Erlangen, Germany), a large-body PET system using approximately 250000 single lutetium oxyorthosilicate (LSO) detector elements covering an axial field of view (FOV) of 106~cm \cite{Prenosil2021_VisionQuadra}. The background radiation in the scanner generated by the presence of $^{176}$Lu in the LSO crystals can be simulated if desired but was neglected in the current simulation. The attenuation map for gamma energies of 511~keV generated from the XCAT phantom was used as an attenuation phantom. The attenuation phantom was positioned in the PET scanner so that the upper body, including the head, was within the FOV of the scanner. Since the simulated cell paths are based on the blood vessel and organ structure of the XCAT phantom, their positions correspond exactly to the anatomical structures in the attenuation phantom, making the simulation also realistic in view of the proportion of attenuated and scattered events. Although the lower body extends beyond the scanner's FOV, it is still simulated to account for cells located there.
The output of the GATE simulations is provided by ROOT files, which can be converted into a format required for the PEPT or PET algorithm applied by the user, e.g. list mode format or sinograms. In this work, no dedicated cell tracking algorithm was applied. Instead, the ROOT output was converted into list mode format and was reconstructed using a standard expectation-maximization algorithm to demonstrate that the simulations yield realistic data.
\section{results}

Starting from the aorta, with a mean blood flow of $100~$ml/s, the blood distributes into all vessels and tissues depending on the resistances of the vessel system and the blood pressure differences (as listed exemplarily in Table~\ref{table}). It is evident that the lungs receive the entire cardiac output through the right ventricular contraction, which accounts for 100~\% of the total blood flow. Notably, the kidneys, gastrointestinal tract, and liver show strong alignment between the simulated flow rate and physiological flow distributions from literature \cite{williams1989reference}, highlighting CeFLoPS's ability to capture predominant systemic blood flow patterns. However, considerable discrepancies are observed for the brain, spleen, pancreas, and thyroid. This behavior can be attributed to the insufficient definition of the vascular structure of these organs or a high vessel resistance due to a large number of small vessels in the case of the brain.
\begin{table}[ht!]
\caption{Blood flow distribution simulated with CeFloPS}
\label{table}
\setlength{\tabcolsep}{3pt}
\centering
\begin{tabular}{|p{70pt}|c|c|}
\hline
Blood vessel or organ & 
\begin{tabular}[c]{@{}c@{}}Simulated blood flow rate\\(\% of total blood flow$^\mathrm{a}$)\end{tabular} &
\begin{tabular}[c]{@{}c@{}c@{}}Reference blood flow rate\\in humans$^\mathrm{b}$\\(\% of total blood flow)\end{tabular} \\
\hline
brain & 
1 &
12 \\
kidneys & 
13 &
19 \\
gastrointestinal tract & 
13 &
15 \\
spleen & 
0.04 &
3.0 \\
pancreas & 
0.09 &
1.0 \\
liver (total) &
24 &
25 \\
muscle &
11 &
17 \\
thyroid &
4 &
1.5 \\
\hline
\multicolumn{2}{p{150pt}}{$^{\mathrm{a}}$For a cardiac output of 100~ml/s.\newline $^\mathrm{b}$Taken from \cite{williams1989reference}.}
\end{tabular}
\label{tab1}
\end{table}

Table~\ref{tab:Reynolds} presents the mean blood flow rate and Reynolds number for the blood vessels in CeFloPS. Vessels are grouped into three size-based categories separately for arteries and veins. As expected from \eqref{eq:average_Q} to \eqref{eq:Reynolds}, large arteries such as the aorta exhibit the highest flow rates and Reynolds numbers, whereas small arteries and veins show markedly lower values, consistent with their higher resistance.
\begin{table}[ht!]
\caption{Average vessel radius ($\overline{r}$), blood flow rate ($\overline{Q}$), and Reynolds number ($\overline{\mathrm{Re}}$) for different vessel types in CeFloPS, differentiated between arteries and veins and further divided into large ($r\geq6\,\mathrm{mm}$), medium ($3\,\mathrm{mm}\leq r<6\,\mathrm{mm}$), and small ($r<3\,\mathrm{mm}$) groups}
\setlength{\tabcolsep}{3pt}
\centering
\begin{tabular}{|p{80pt}|c|c|c|}
\hline
Vessel type & 
\rule[-1.5ex]{0pt}{4ex} $\overline{r}$ [mm] &
\rule[-1.5ex]{0pt}{4ex} $\overline{Q}$ [ml/s] &
\rule[-1.5ex]{0pt}{4ex} $\overline{\mathrm{Re}}$\\
\hline
\begin{tabular}[c]{@{}l@{}}large arteries\\(e.g. aorta)\end{tabular} & 
8.49 &
70.69 &
1568 \\
\begin{tabular}[c]{@{}l@{}}medium arteries\\(e.g. arteriae femorales)\end{tabular} &
3.97 &
16.03 &
685 \\
\begin{tabular}[c]{@{}l@{}}small arteries\\(e.g. arteriae renales)\end{tabular} & 
0.81 &
0.39 &
58 \\
\begin{tabular}[c]{@{}l@{}}large veins\\(e.g. venae cavae)\end{tabular} & 
6.51 &
35.12 &
1041 \\
\begin{tabular}[c]{@{}l@{}}medium veins\\(e.g. venae hepaticae)\end{tabular} & 
3.92 &
7.24 &
357\\
\begin{tabular}[c]{@{}l@{}}small veins\\(e.g. sinus sigmoidei)\end{tabular} & 
0.96 &
0.59 &
82\\
\hline
\end{tabular}
\label{tab:Reynolds}
\end{table}

\begin{figure}[ht!]
\centerline{\fbox{\includegraphics[width=.5\columnwidth]{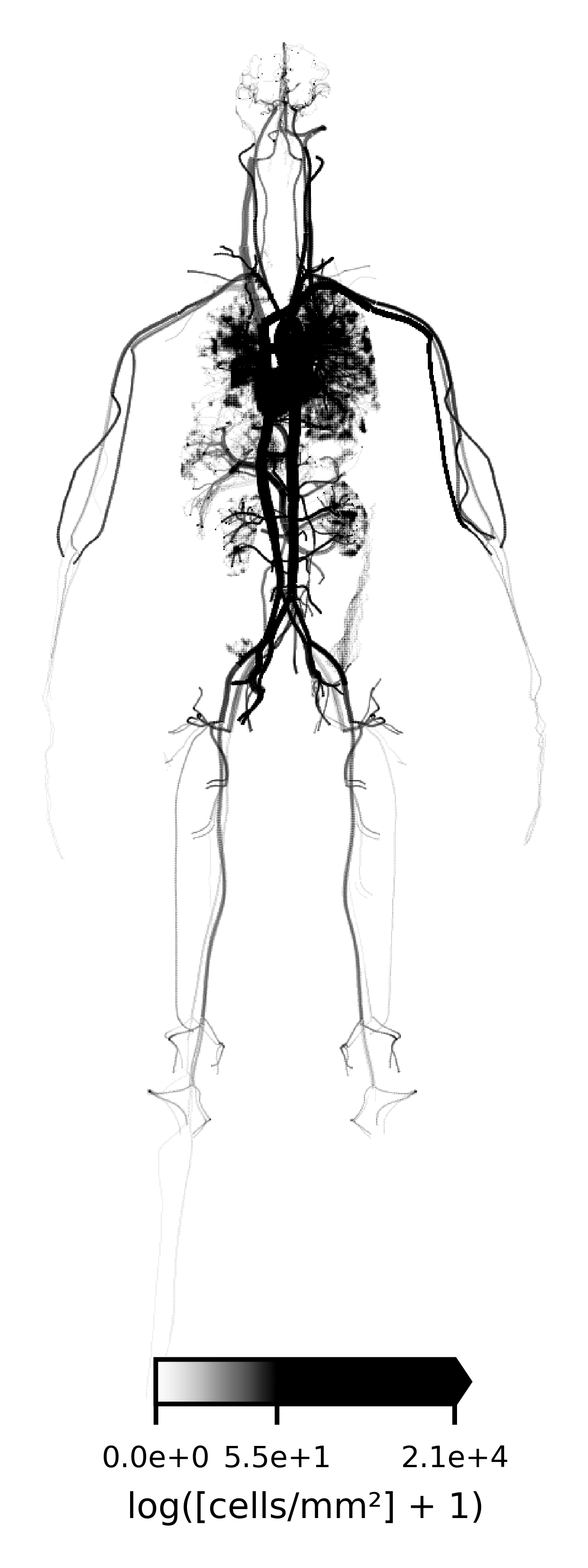}}}
\caption{Maximum intensity projection (1.5~mm pixel size) of the vessels and tissues traversed by 100,000 cells injected into the right arm vein during the first 100 seconds of a simulation. 
The gray log scale encodes the number of cells per mm$^2$.}
\label{body_MIP}
\end{figure}

Fig.~\ref{body_MIP} depicts a maximum intensity projection (MIP) of the distribution of 100,000 injected cells over the first 100 seconds following injection into the left arm vein. As would be anticipated, the route from the injection site to the heart, the heart itself, the pulmonary arteries and veins, the aorta and arteries close to the aorta are traversed with high frequency. In regions such as the legs and arms, it is evident that certain arteries and veins have yet to be reached by a significant number of cells. Additionally, there are observable transitions of cells into organs and backflow into veins, particularly evident in the kidneys, liver, and vena cava. 

\begin{figure}[ht!]
\centerline{\includegraphics[width=\columnwidth]{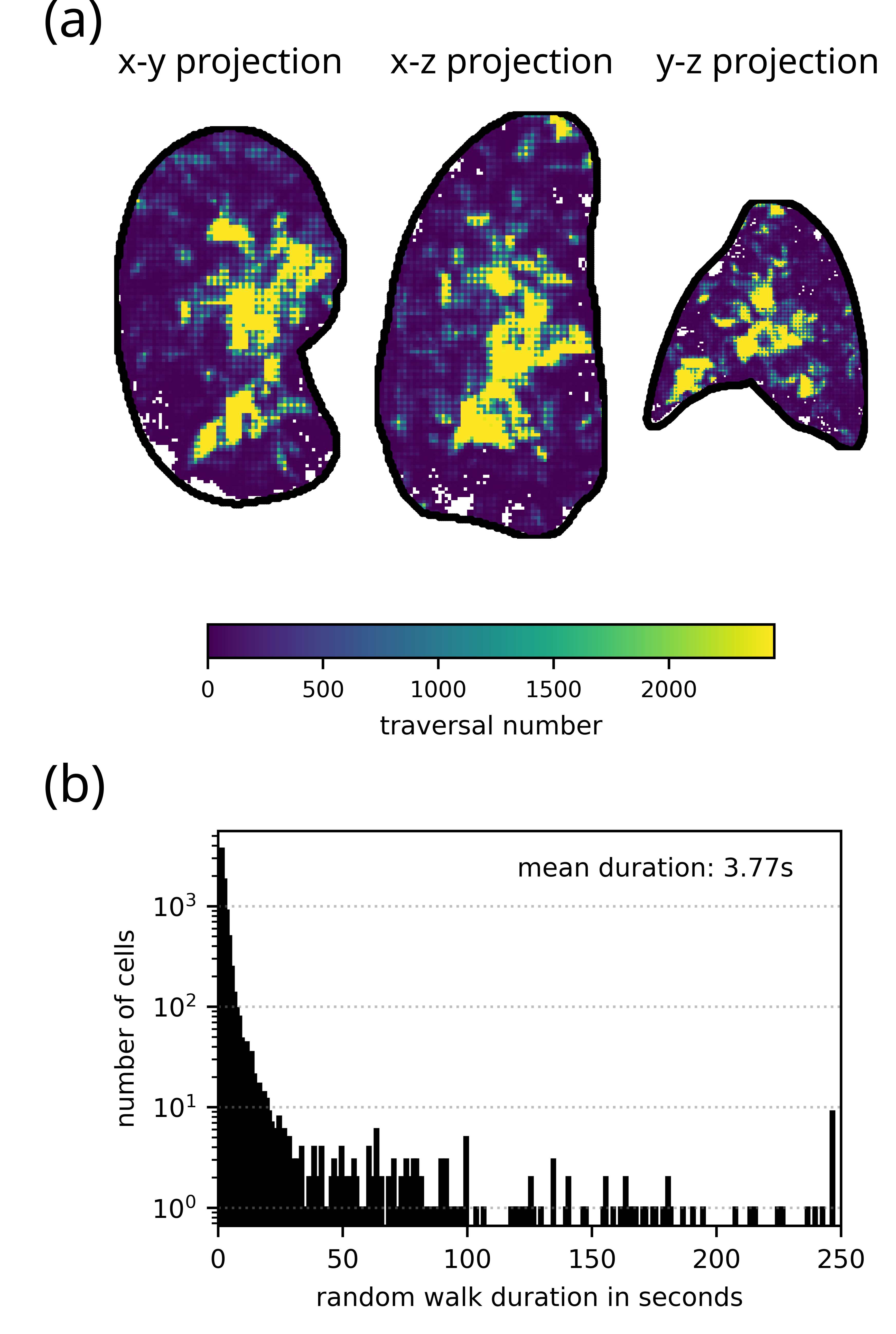}}
\caption{(a) Results from a 3D biased random walk simulation of cell motion within the right lung over a period of 2000~s, visualized as maximum intensity projections (MIPs) in three orthogonal planes. Heatmap intensities indicate voxel traversal frequencies by cells. The cell entry distribution is proportional to simulated blood flow rates at arterial sources. (b) Histogram of random walk durations, illustrating the time required for cells to traverse from arterial sources to venous sinks, with a mean duration of 3.77~s.}
\label{3D_random_walk_lung}
\end{figure}

The biased 3D random walk that cells perform when they migrate from the blood into a tissue is illustrated in Fig.~\ref{3D_random_walk_lung}a using the example of the lung for a duration of 2000~s. The simulation outcome is visualized through MIPs in three orthogonal planes to clearly illustrate spatial traversal patterns. The distribution of cells entering the random walk through the arterial sources is weighted by simulated arterial blood flow fractions of the vessels in the lung. It prominently demonstrates regions of enhanced cellular traversal intensity, indicative of high-flow vascular pathways, and shows that the biased random walk ensures that the entire organ tissue is potentially accessible by incoming cells. Furthermore, the temporal dynamics of cell migration are elucidated in Fig.~\ref{3D_random_walk_lung}b, where the histogram of random walk durations reveals a rapid traversal of the lung vasculature for the majority of cells. The mean duration is 3.77~s, consistent with the theoretical specification outlined in \cite{leggett1995proposed}.

\begin{figure*}[ht!]
\centerline{\includegraphics[width=\textwidth]{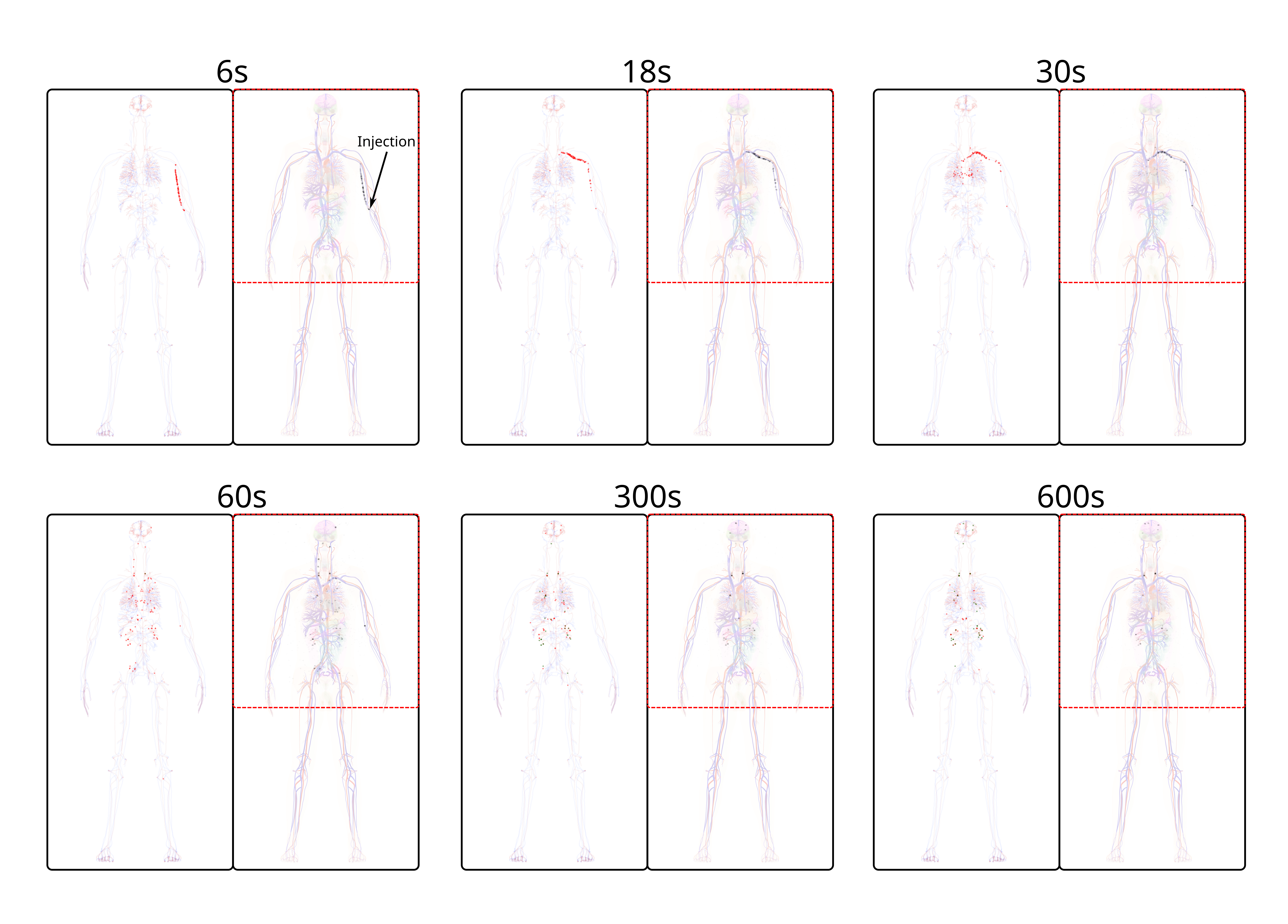}}
\caption{Output of the flow model (left) showing simulated ground-truth cell positions for 100 cells at six different time points, including both moving cells (red) in the node system as well as stationary cells that have entered $C_1$ or $C_2$ (green). For illustration, PET maximum intensity projections (right) using a standard EM based image reconstruction show the cell distribution (black) superimposed on the organ and blood vessel structure. The red dotted line marks the FOV of the simulated PET scanner.}
\label{Distribution}
\end{figure*}

Fig.~\ref{Distribution} illustrates the locations of 100 simulated cells within a node system at six time points, corresponding to the output of the flow model, compared to MIPs reconstructed from PET data using an expectation-maximization based image reconstruction without attenuation correction. The reconstructed positions and cell counts align well with ground-truth positions, despite minor reconstruction artifacts causing varying intensities in the MIPs due to missing attenuation and scatter correction. Initially, cells (red spheres) are injected into the left arm vein (median cubital) over ten seconds. After six seconds, the first cells migrate up the vein while new cells are still injected. By 18 to 30 seconds, cells have reached the left subclavian vein and heart and start to move into the lungs. At 60 seconds, the majority of cells are located in vasculature and organs of the thorax, with visible spread towards the lungs and kidneys. A slowly traveling cell remaining near the injection site can be observed. Over longer timescales (300–600~s), the distribution extends across the torso, with cells accumulating in the organs and transitioning to compartments $C_1$ or $C_2$, changing color from red to green. In the MIP, this leads to spots of higher intensity, most notably visible in the kidneys, thyroids, and the brain.

\begin{figure}[ht!]
\centerline{\includegraphics[width=\columnwidth]{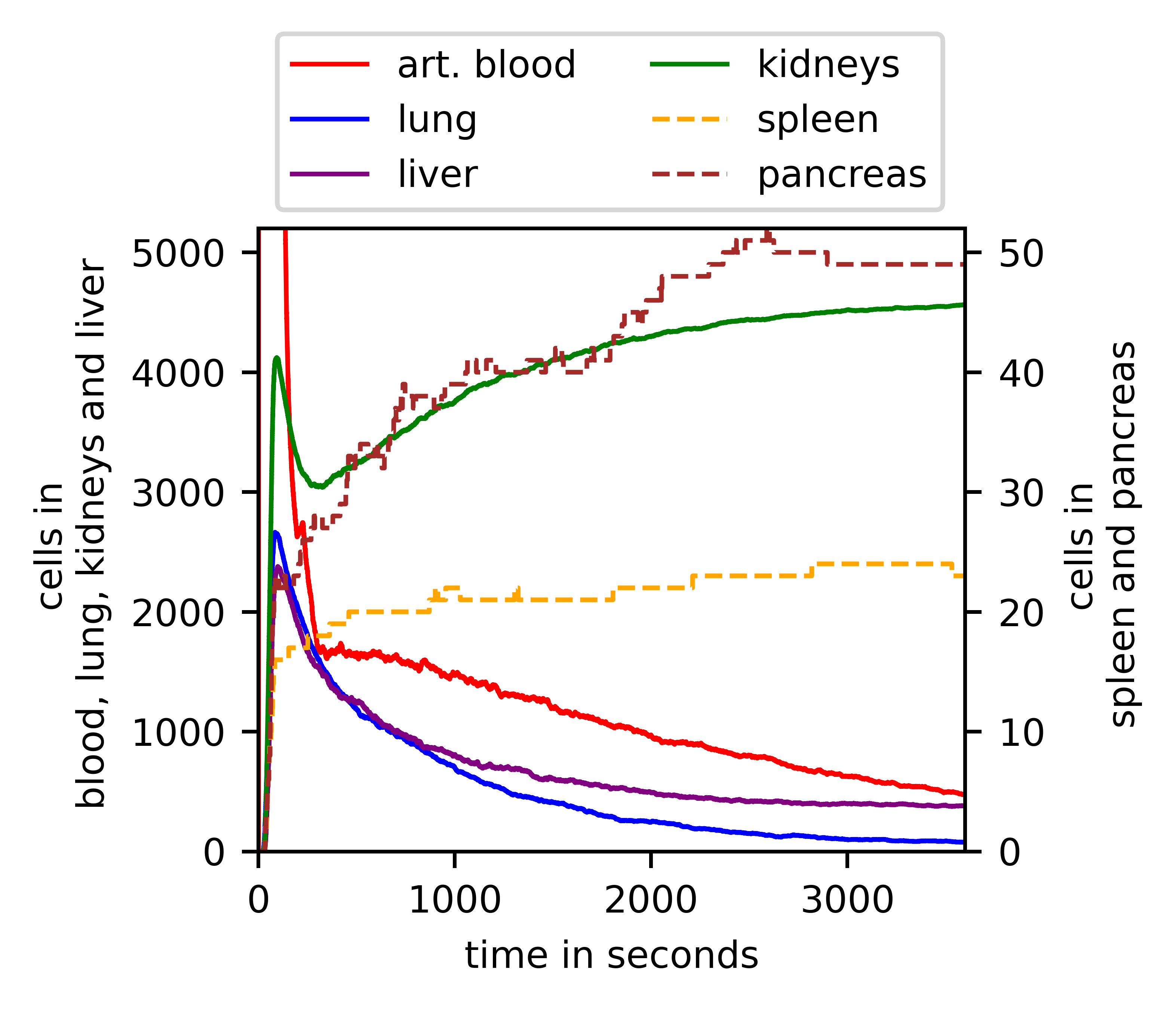}}
\caption{Temporal distribution of simulated number of cells across arterial blood and specific tissues. Left y-axis corresponds to arterial blood, lung, liver, and kidneys, whereas right y-axis denotes cell counts in spleen and pancreas due to differing scale ranges.}
\label{cellnumber_init}
\end{figure}

\begin{figure}[ht!]
\centerline{\includegraphics[width=\columnwidth]{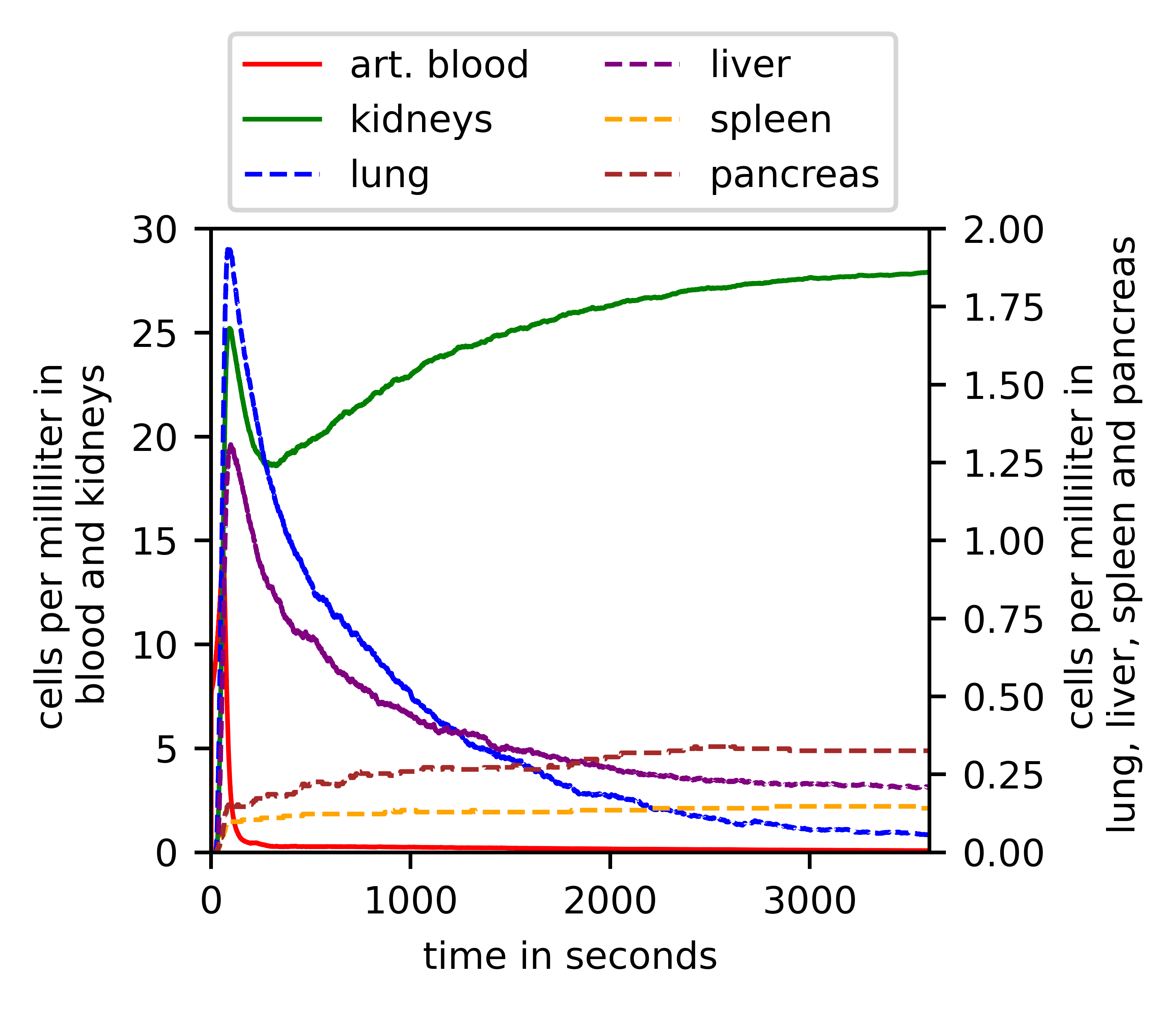}}
\caption{Temporal distribution of simulated cell concentrations (cells per milliliter) across arterial blood and specific tissues. Left y-axis represents arterial blood and kidneys; right y-axis depicts concentration in lung, liver, spleen, and pancreas to highlight scale discrepancies.}
\label{concentration_init}
\end{figure}

Figs.~\ref{cellnumber_init} and \ref{concentration_init} illustrate the temporal distribution of 100,000 cells in arterial blood, lung, kidneys, liver, spleen, and pancreas as absolute cell numbers (sum of cells in $C_1$ and $C_2$) and as cell concentrations per milliliter tissue, respectively. Initially, the cell count in the arterial blood rapidly drops and a pronounced peak of cell presence is observed in the lung, liver, and kidneys, indicative of early vascular dissemination and transitions from $C_A$ to $C_1$. Subsequently, cell counts decline within these tissues, aligning with physiological expectations of cells transitioning back from $C_1$ to $C_A$ and leaving the tissue into the venous blood. Notably, over extended simulation time, the cell count in the kidneys rises again and the kidneys and the pancreas exhibit a progressive increase in cell accumulation, indicating trapping of cells in $C_2$. Meanwhile, the spleen maintains a relatively steady cell population, reflecting a possible equilibrium between cell influx and clearance. The concentration plots (Fig.~\ref{concentration_init}) underscore these observations by correcting for organ volume, thus offering additional insight into organ-specific cell affinity beyond raw cell counts. In terms of raw cell count, spleen and pancreas reach significantly lower numbers than the other tissues, which is not apparent in the concentration plot. Notably, the kidneys exhibit considerably higher values than other tissues with regard to both the absolute number of cells and the cell concentration.


\begin{figure}[ht!]
\centerline{\includegraphics[width=\columnwidth]{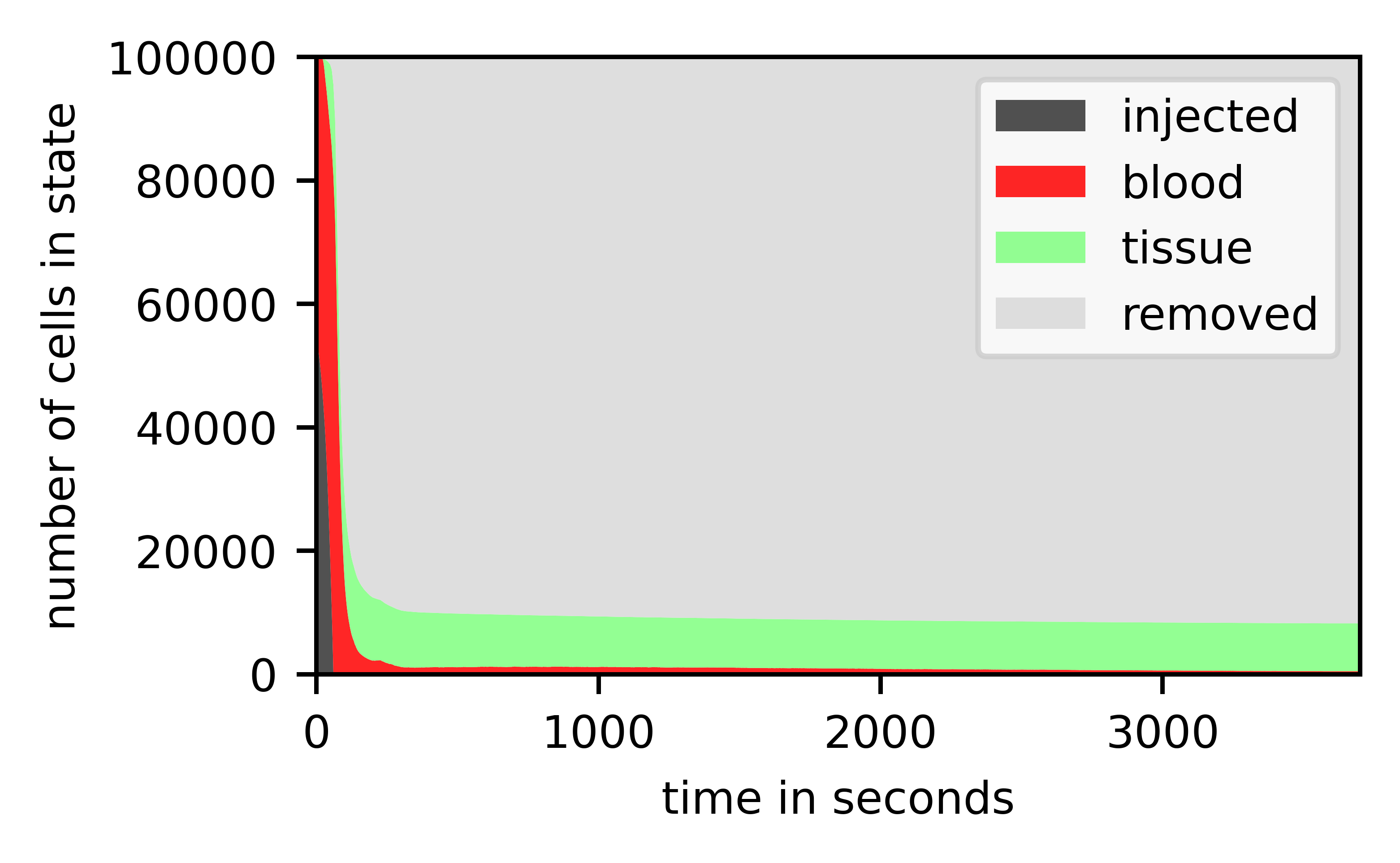}}
\caption{Temporal distribution of 100,000 simulated cells across compartments shown in Fig.~\ref{Compartments}: injected, arterial blood, tissue, and removed.}
\label{removed_cells}
\end{figure}

Fig.~\ref{removed_cells} illustrates the temporal evolution of the 100,000 simulated cells across the virtual, blood, and tissue compartments depicted in Fig.~\ref{Compartments} and described by \eqref{AIF_equation}. After the 30~s injection period, all cells have transitioned from $C_\mathrm{inj}$ to $C_A$. This is followed by a swift redistribution into tissues or removal from blood and transition into $C_\mathrm{rem}$. After a brief transient phase, a stable equilibrium emerges, characterized predominantly with over 80\% of cells in the removed state. The dynamics clearly show that only a small fraction of injected cells persist within blood or the defined tissue compartments at longer time points. The majority of cells are located in tissues without specified dynamics, like skin, muscle or adipose. 

\begin{figure*}[ht!]
\centerline{\includegraphics[width=\textwidth]{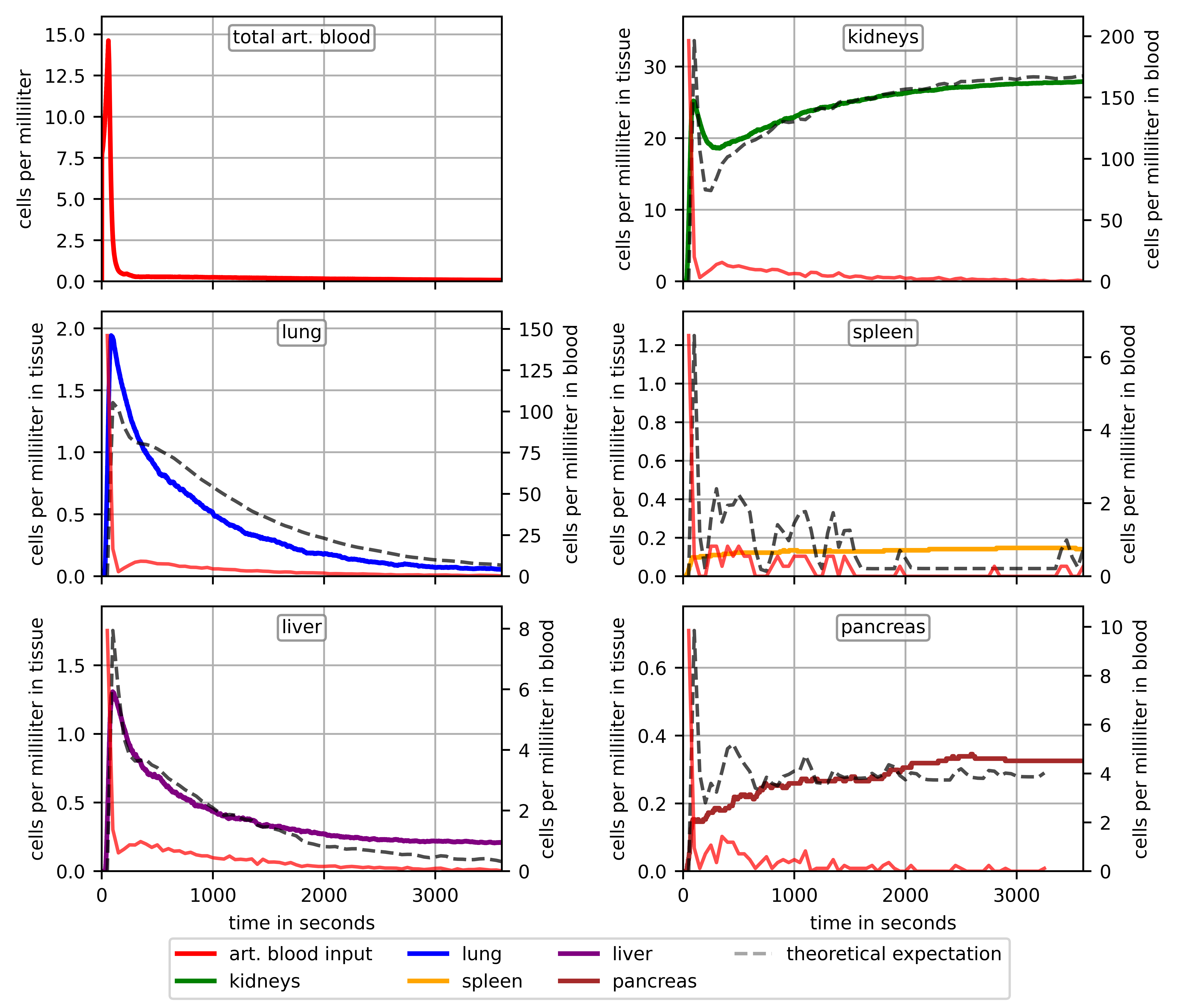}}
\caption{Simulated concentration distributions profiles of cells in tissues and tissue specific arterial blood input. Solid colored lines represent the simulated cell concentrations per milliliter in the respective tissues, whereas the red lines show the cell concentration in the arterial blood perfusing each tissue. The plot in the upper left displays the total arterial blood concentration in the whole body. The dashed black lines depict the theoretically expected concentration distributions derived from \eqref{diffequations_two}, based on the tissue specific arterial input functions and transition rates (Table \ref{tab:k_values}).}
\label{Comparison}
\end{figure*}

Fig.~\ref{Comparison} compares the simulated cell concentrations in the tissues to the distributions one would expect from \eqref{diffequations_two}. To calculate the theoretically expected distributions for each tissue, the number of cells perfusing from the arterial blood into tissue are mapped. The arterial input to each tissue is determined by the unique and locally dependent distribution of blood flow and the subsequent delay and dispersion effects. Together with the transition rates (Table \ref{tab:k_values}) of each tissue, the arterial input allows derivation of the expected cell distributions from the differential equations. The reference volume to which the cells of the arterial input function are normalized in order to obtain a concentration is treated as a tissue-specific variable. This variable is dependent on the volume of the small arteries, arterioles and capillaries leading to and perfusing the tissue. This factor was determined by fitting the two distributions and multiplying the result with the arterial input function. This yields the concentration of cells in the arterial blood and the expected theoretical curve. As shown in fig.~\ref{Comparison} the simulated distributions describe the theoretically expected ones well, with small over- or undershoots of the initial peaks. In the case of the spleen and pancreas, the theoretical expected cell concentration exhibits fluctuations, due to the small number of cells in these tissues.

The performance of the simulation is linked to the parameters chosen to prepare the data as well as parameters regarding a desired output. We employed multiprocessing with python and launch multiple instances, each simulating a fraction of all cells. By default, no synchronization between those processes is necessary, as only positional data are read from shared memory. A new implementation making use of disabled Global Interpreter Lock (GIL) further reduces runtime. The runtime generally scales linearly with the number of cells simulated. Scaling with longer simulated time is less than linear, as cells in compartment $C_2$ are removed from the simulation after storing their location. The process of data preparation on the other hand also depends on specified parameters; here the resolution of pipe volumes that vessels are separated into, voxel sizes and desired distance between reference nodes inside of vessel meshes play a crucial role. In this step the flow fields are also calculated using a CUDA kernel. Once the data preparation is finished, it is stored and can be loaded for consecutive simulations. Our simulation of 100,000~cells over a duration of 3600 seconds required about 45~minutes on an AMD EPYC 8324P 32-Core Processor (64 threads) using 32 spawned processes. The data preparation beforehand took approximately 1.5 hours.
\section{discussion}

The simulation framework presented here, CeFloPS, has been demonstrated to be capable of simulating physiologically meaningful cell pathways within the human body, which can be utilized for the development of image reconstruction algorithms as well as specific cell tracking strategies. CeFloPS allows to simulate not only the flow of cells through the vessel branches but also the migration of cells from the vasculature into tissues either as reversible or non-reversible trapping steps. In this way, different physiologically relevant scenarios can be simulated for various cell types providing viable data for mimicking the dynamic distribution of cells within the human body. Beyond pure cell-based simulations, even tracer studies could be simulated by CeFloPS, as shown here by the FDG example. Following the CeFloPS simulations, a label can be assigned to the cells, which was chosen for PET studies as a radio-label ($\beta^+$ emitter) but could be of any kind allowing also simulations for other imaging modalities, such as SPECT, OI, MRI, or ultrasound imaging. 

The cell flow simulation presented is demonstrated and partially validated using FDG as a model for "cell" motion. As FDG is a well known tracer for PET which provides reference values from numerous publications, organ-specific compartment model parameters and resulting tracer concentrations could be utilized for our study. Such information is currently hardly available from cell tracking studies making translation towards specific cell types (e.g. T-cells, immune cells, etc.) difficult to develop and validate. As specific data is available for specific cell flow behavior in humans, dedicated CeFloPS simulations can be developed. Comparing the MIP of the cell distribution in the first 100 seconds (Fig.~\ref{body_MIP}) and the simulated position of 100 cells over 600 seconds (Fig.~\ref{Distribution}) to similar MIPs of dynamic total-body FDG PET scans like \cite{liu2021kinetic} and \cite{liu2021ultra}, a similar flow behavior can be observed. Both the cells in the simulations and the FDG in the PET scans take approximately 20 to 30 seconds to traverse the distance from the injection site (arm or leg, respectively) to the heart and lungs. The PET scans show FDG accumulation in the kidneys after 35 to 40 seconds, which subsequently maintain a high signal over the entire scan time. In the simulations, the cell accumulation in the kidneys is observable after approximately 60 seconds and persists until the maximum simulation time of 10 minutes. Large blood vessels can still be distinctly identified in the scans after 10 minutes due to their signal intensity. This is also consistent with the simulations, where after 10 minutes a significant proportion of the cells are located in the vascular system, as can be seen from the color coding in Fig.~\ref{Distribution}.

The simulation of PET physics was performed by GATE using a large-body PET system where radioactivity was attributed to each individual cell, defined as a small ball. Since this post-processing step is independent of the CeFloPS simulation, any other simulation framework could also be used which is capable of processing dynamic source distributions. In this study, only one bed position of 106~cm length defined by a large-body PET was simulated leading to an incomplete coverage of the entire human body. However, this could easily be extended towards a multi-bed simulation or using a total-body PET system covering a larger FOV.

The radioactivity of 1~kBq attributed to each FDG ``cell" in the example is quite high and not realistic in the context of cell tracking experiments. In any case, this served only as a proof-of-principle value allowing a relatively noise-free dynamic image reconstruction of the simulated PET data for comparison with ground-truth cell positions. For realistic cell tracking simulation, radioactivity values in the range of a few Bq would be more appropriate. However, the demonstration of cell trajectory based PET reconstructions for low-count data, as shown before by our group (see \cite{Schmitzer2019_OptimalTransport}), was not part of this work and will be the subject of subsequent studies.

It is evident from the cell numbers (Fig.~\ref{cellnumber_init})  that certain organs are preferentially traversed by cells over others. This is due to the fact that the probability of which blood vessel branch is chosen at bifurcations is dependent upon the proportion of blood flow, as illustrated by \eqref{eq:link_probability}. Organs that are less likely to be traversed by cells will therefore have a lower cell count. This is due to the rudimentary modeling of the blood vessels in the kidneys in XCAT, exhibiting a lack of small diameter vessels (see Fig.~\ref{XCAT}c). Given that these represent a significant resistance, as indicated by \eqref{eq:vessel_resistance}, the blood flow to the kidneys is currently overestimated. Consequently, the probability of cell paths traversing the kidneys is greater than that of other organs. The spatial resolution of the XCAT phantom thus represents a clear limitation of CeFloPS. In order to improve towards a more realistic behavior, two potential approaches could be followed. Either, missing smaller blood vessels could be added to the XCAT phantom which would lead to a higher resistance in the vasculature and thus to a more realistic blood flow. A potential approach is illustrated in \cite{8094255}, which outlines techniques for extending the pulmonary vasculature in the lungs. Or, pre-defined resistances could be attributed to dedicated organs which would also lead to a more realistic flow simulation in CeFloPS. In principle, a different anthropomorphic phantom with higher level of details could also be implemented in the CeFloPS framework which could circumvent the flow imbalances.

Blood flow is currently implemented as a continuous laminar flow, completely ignoring the pulsating characteristics of the blood circulation. The assumption of laminar flow in the blood vessels is a highly simplified model that allows for rapid determination of cell positions within the vessels without great computational effort and is supported by the Reynolds numbers in the blood vessels, which were obtained for all vessels. All Reynolds numbers are well below the threshold value of 2000 at which transition to turbulent flow would occur \cite{CDT9829}.  For example, the Reynolds numbers of the aorta range from 1300 to 1900. In addition, no pulsatile blood flow is allowed in this state of the simulation which further supports our assumption. A more comprehensive depiction of blood flow would encompass pulsed flow, along with additional parameters such as stress in arterial walls or hemorheology and would require solving the Navier-Stokes equation, as outlined for example in \cite{skalak1989} and \cite{secomb2016}. Other approaches involve the utilization of finite element or Lattice Boltzmann methods to model blood flow within the vessels, the latter having previously been employed for the arterial vessels of the XCAT phantom, as demonstrated in \cite{10.1145/2807591.2807676}. With regard to the intended field of application of CeFloPS, namely the generation of anatomically and physiologically meaningful cell pathways for PET simulations, the computational effort of these methods exceeds the scope of our study. Nevertheless, these methods could be of interest when focusing on pulsed flow and local effects due to the pulse wave behavior. Also, turbulent flow and a variable blood viscosity, which is known to play an important role in the area of the capillary bed, are completely ignored. Moreover, cells in our study are considered as small particles, which freely flow through the blood system and freely diffuse through the tissue compartments. Cell-specific behaviors, such as rolling or adhesion at the endothelium or immune related activation cannot explicitly be modeled within the CeFloPS simulation. However, a similar behavior could potentially be modeled by setting (dynamic) $k$-parameters (Tab.~\ref{tab:k_values}) for compartmental exchanges.
\section{conclusion}

In this study, we presented the development of a cell flow PET simulation (CeFloPS) that is capable of simulating the behavior of injected radio-labeled cells during human whole-body PET imaging with ground-truth knowledge of the cells' positions. The data obtained with CeFloPS is of significant value for the development and validation of cell tracking algorithms, with the potential to contribute to a deeper understanding of immune system-related processes or to the guidance of cell-based therapies.
\section*{Code availability}
The complete CeFloPS code (with the exception of the XCAT-based StL files) is available on GitHub: \url{https://github.com/EIMI-Institute/CeFloPS} (DOI: \url{https://doi.org/10.5281/zenodo.17100367}).
\section*{Acknowledgment}
We greatly acknowledge the help of Paul Segars for providing the STEP files of the XCAT phantom with precise vessel and tissue definitions.

\bibliography{99_bibliography}

\end{document}